\documentstyle[12pt,aasms4]{article}

\lefthead{Boggs et al.}
\righthead{Diffuse Galactic Soft Gamma-Ray Emission}

\begin{document}

\title{Diffuse Galactic Soft Gamma-Ray Emission}

\author{S. E. Boggs\altaffilmark{1,2}, R. P. Lin\altaffilmark{2}, and
    S. Slassi-Sennou}
\affil{Space Sciences Laboratory, University of California,
    Berkeley, CA 94720}

\and

\author{W. Coburn and R. M. Pelling}
\affil{Center for Astrophysics and Space Sciences, University of
    California, La Jolla, CA 92093}

\altaffiltext{1}{Millikan Postdoctoral Research Fellow;\\
    present address: Space Radiation Laboratory,
    California Institute of Technology, MC 220-47, Pasadena, CA 91125;
    boggs@srl.caltech.edu}
\altaffiltext{2}{Department of Physics, University of California,
    Berkeley, CA 94720}

\begin{abstract}
The Galactic diffuse soft gamma-ray (30-800 keV) emission
has been measured from the Galactic Center by the 
HIREGS balloon-borne germanium spectrometer to determine
the spectral characteristics and origin of the emission. The 
resulting Galactic diffuse continuum is found to agree well 
with a single power-law (plus positronium) over the entire 
energy range, consistent with RXTE and COMPTEL/CGRO
observations at lower and higher energies, respectively.
We find no evidence of spectral steepening below 200 keV,
as has been reported in previous observations.
The spatial distribution along the Galactic ridge is 
found to be nearly flat, with upper limits set on the
longitudinal gradient, and with no evidence of an edge in the 
observed region.
The soft gamma-ray diffuse spectrum is well modeled by inverse
Compton scattering of interstellar radiation off of cosmic-ray
electrons, minimizing the need to invoke inefficient
nonthermal bremsstrahlung emission. The resulting power requirement
is well within that provided by Galactic supernovae.
We speculate that the measured spectrum provides the first direct
constraints on the cosmic-ray electron spectrum below 300 MeV.
\end{abstract}


\section{Introduction}

Galactic diffuse gamma-ray emission provides information 
on the origin, propagation, and interaction of cosmic-ray 
electrons within the Galaxy, complementary to the knowledge
gained from direct observation of the cosmic-ray
electrons in the Solar vicinity, and nonthermal radio emission 
from interactions with Galactic magnetic fields. The
cosmic-ray electron spectrum has been directly measured above 
a few GeV (\cite{gol84,gol94,tai93}), and 
extended down to 300 MeV by radio synchrotron observations 
(\cite{web83}). Below 300 MeV, the electron spectrum 
must be constrained by the gamma-ray observations
themselves. Observations and mapping of this diffuse continuum 
also provides a background which, as soft gamma-ray 
observations become more sensitive, will be important for 
the detailed study of faint sources.

A number of satellite and balloon instruments have measured
the Galactic diffuse continuum in the hard X-ray/soft 
gamma-ray bands, e.g., OSO 7 [7-40 keV] (\cite{whe76}), 
HEAO 1 [2-50 keV] (\cite{wor82}), SMM [0.3-8.5 MeV] 
(\cite{har90}), GRIS [20 keV - 10 MeV] (\cite{geh93}),
COMPTEL/CGRO [1-30 MeV] (\cite{str96}),
Ginga [2-16 keV] (\cite{yam96}),
OSSE/CGRO [70 keV - 4 MeV] (\cite{pur95}),
RXTE [3-35 keV] (\cite{val98}). These observations have 
revealed nonthermal emission distributed along a Galactic 
ridge that extends $\pm 60^{\circ}$ in longitude.

The hard X-ray/soft gamma-ray continuum is believed to be 
dominated by three components (\cite{str95,ski93}):
nonthermal bremsstrahlung from
cosmic-ray interactions with the interstellar gas, inverse 
Compton scattering of the interstellar radiation (optical, 
infrared, and cosmic microwave background) off of the
cosmic-ray electrons, and positronium continuum from the 
annihilation of Galactic positrons below 511 keV. At lower 
energies ($<10 keV$) thermal emission from the hot ISM 
dominates, while at higher energies ($>100 MeV$) $\pi^{0}$ decay 
emission dominates.

Models of the nonthermal bremsstrahlung and inverse 
Compton components (\cite{str95,ski93}) predict
that bremsstrahlung dominates in the 
MeV range, though inverse Compton still provides a
significant fraction of the total flux; at lower energies, however, 
bremsstrahlung is highly inefficient compared to ionization 
and Coulomb collision losses (e.g., \cite{str95,ski96}),
and the bremsstrahlung 
spectrum drops rapidly; therefore, inverse Compton is 
expected to dominate below 1 MeV. The combination of the 
bremsstrahlung and inverse Compton, however, should produce
a very smooth continuum from 10 keV to 30 MeV 
(\cite{str95,ski93}), though the 
addition of the positronium continuum will produce a
well-defined spectral jump at 511 keV. This smooth continuum 
makes it difficult to distinguish the 
relative contributions of the bremsstrahlung and inverse 
Compton emission from the spectrum alone.

The spatial distributions of the two components, however, 
should be significantly different. Due to the diffuse cosmic 
background, and the large scale heights of the interstellar 
optical and IR emission and the cosmic-ray electrons compared
to the interstellar gas, the scale height of the inverse 
Compton component should be significantly larger than the 
scale height of the bremsstrahlung emission. The presence 
of both broad and narrow scale height components has been 
confirmed in the 1-30 MeV energy band by COMPTEL/CGRO
(\cite{str96}) and in the 3-35 keV band by 
RXTE (\cite{val98}), adding strong support 
to the two-component model. These 
observations also suggest a smooth continuum from 10 keV 
to 30 MeV, with measured photon indices of 1.8 (RXTE) 
and $\sim$2.0 (COMPTEL/CGRO). This smooth continuum is 
consistent with models of the combined bremsstrahlung and 
inverse Compton emission (e.g., \cite{str95,ski93}).

Observations in the 50 keV - 4 MeV energy band with 
OSSE/CGRO, the energy range between COMPTEL/CGRO
and RXTE, have brought into question this simple 
two component (plus positronium) model of the Galactic 
diffuse continuum (\cite{pur95}). These observations 
suggest a strong steepening of the diffuse spectrum below 
200 keV which is difficult to explain in terms of this 
model. The inverse Compton emission at these energies is 
produced predominantly by higher energy ($>GeV$) electrons,
the spectrum of which has been well constrained by 
Galactic synchrotron radio emission (\cite{web83}), so that 
an inverse Compton origin for a hard X-ray excess is highly 
unlikely (\cite{ski97}). Furthermore, 
due to the inefficiency of bremsstrahlung emission below 
1 MeV, the energetics required to produce such hard X-ray 
emission exceeds the total power expected from Galactic 
supernovae (\cite{ski97}).

Measurements of the diffuse Galactic hard X-ray and soft
gamma-ray emission are inherently upper limits due to the large 
number of hard X-ray compact sources in the Galactic Center
region and the inevitable source confusion. Difficulty 
arises because non-imaging instruments with large
fields-of-view (FOVs) are most sensitive to the diffuse continuum 
but are highly susceptible to compact source confusion, 
whereas imaging instruments or instruments with narrow 
FOVs are able to measure the compact sources individually, 
but are not as sensitive to the diffuse emission. Scans by 
wide and moderate FOV instruments (e.g., HEAO 1, SMM), and coordinated
observations between imaging instruments and wider FOV
instruments (e.g., OSSE/CGRO \& SIGMA) 
have been performed to subtract out the compact source 
contributions, but still show hard X-ray spectra that are
easiest to explain by compact source confusion.

The HIREGS spectrometer observed the Galactic Center 
region in January 1995 during a long duration balloon flight 
from Antarctica (\cite{bog98}). HIREGS is a wide 
FOV instrument ($21^{\circ}$ FWHM), but three sets of narrow 
FOV passive hard X-ray collimators ($3.7^{\circ} \times 21^{\circ}$ FWHM) 
were added over half of the twelve detectors for this flight. 
HIREGS thus simultaneously performed wide and narrow 
FOV observations of the same region, allowing the direct 
measurement and separation of compact source spectra and 
the Galactic diffuse.

In this paper we present the analysis and results of the 
HIREGS measurements of the diffuse Galactic emission in 
the 30-800 keV band from the Galactic Center region. Our 
observations are consistent with a single power-law (plus 
positronium) over the entire energy band, with a spectral 
index that agrees well with RXTE observations at lower 
energies, and COMPTEL/CGRO observations at higher energies.
These results are inconsistent with previous observations
of spectral steepening below 200 keV (e.g., OSSE/CGRO),
which are most likely due to compact source contamination.
Upper limits on the longitudinal gradients along 
the Galactic ridge are presented. We discuss implications
of the soft gamma-ray diffuse Galactic emission on the
cosmic-ray electron spectrum.

\section{Instrument and Observations}

HIREGS is a high-resolution spectrometer (2.0 keV 
FWHM at 592 keV) covering three orders of magnitude in 
energy, from 20 keV to 17 MeV. The instrument is an array 
of twelve high-purity 6.7-cm diameter by 6.1-cm height 
coaxial germanium detectors designed for long duration 
balloon flights. The array is enclosed on the sides and bottom
by a 5-cm thick BGO shield, and collimated to a $21^{\circ}$ 
FWHM FOV by a 10-cm thick CsI collimator. The instrument
(\cite{pel92}), and the flight (\cite{bog98}) 
are described in greater detail elsewhere.

Passive molybdenum (0.15 mm) collimators were designed 
for this flight to limit the hard X-ray ($<200 keV$) FOV for 
half of the detectors to $3.7^{\circ} \times 21^{\circ}$ FWHM
(Figure 1). Three sets of these collimators were inserted
in the main collimator and slanted in parallel directions, 
providing three parallel but separate FOVs.
Therefore, the instrument
had four different hard X-ray FOVs for any given 
pointing: one wide $21^{\circ}$ FOV, and three narrow
$3.7^{\circ} \times 21^{\circ}$ FOVs.

Spectra were collected for 30 minute periods, alternating 
between source observations and background observations 
at the same horizon elevation. The backgrounds were taken 
at least one FOV width off of the Galactic Plane to avoid 
source contamination, and were alternated to the east and 
west of the source pointings to avoid any systematic errors. 
Despite these precautions, there is a soft source ($<40 keV$) 
contaminating the westerly background pointings. This 
source affected the preliminary analysis of this data
(\cite{bog97}), allowing misidentification of the compact 
sources and forcing one spectra to show significant spectral 
flattening below 40 keV. For this analysis, the westerly 
background spectra and corresponding source spectra are 
not used for energy bins below 40 keV.

The data reported here are for three different pointing
directions (Figure 1): 48 hrs centered on the Galactic Center 
(l=$0^{\circ}$, b=$0^{\circ}$), 4 hrs on GRO J1655-40
(l=$-14.6^{\circ}$, b=$1.8^{\circ}$), 
and 23 hrs on the Galactic Plane
(l=$-25^{\circ}$, b=$0^{\circ}$).


\section{Conversion to Photon Spectra}

The source and background counts were gain and livetime 
corrected, and then binned into 42 channels from 20 keV to 
17 MeV for each separate FOV. A narrow interval around 
the neutron-induced instrumental line at 198 keV was 
excluded due to its variability, and the 511 keV line
subtraction was performed in a separate analysis (\cite{phd98}). 
The rest of the lines were either weak enough or varied 
slowly enough that they required no special treatment.

The corresponding background is subtracted from each 
source observation and the residual count spectra are
corrected for atmospheric attenuation and detector response by 
a procedure equivalent to directly inverting the coupled 
atmospheric/detector response matrix. Matrices were
modeled for a range of atmospheric column densities using a 
Monte Carlo simulation of a thin atmosphere above the 
HIREGS instrument, then normalized to calibration
measurements performed before the flight (\cite{phd98}). For 
the exact column density of any source observation, the 
response matrix is interpolated from the models.

All of the photon spectra for a given pointing direction are 
then weighted and averaged into a single spectrum for each 
FOV. These spectra are weighted by the product of livetime 
and atmospheric transmission, which is an energy-dependent
factor that is proportional to the expected signal-to-noise.
Spectra are weighted by this factor instead of the 
measured signal-to-noise so as not to bias the average 
towards those spectra where statistical fluctuations have 
resulted in a higher signal.

\section{Determination of Compact Sources}

The hard X-ray (30-200 keV) photon flux for each of the 12 
FOVs is shown in Figure 2. The flux has been corrected for 
the detector response and the atmospheric transmission, but 
not for the angular response of the main collimator or the 
narrow collimators. Therefore, these rates are the sum of 
the diffuse and compact sources in the Galactic Center 
region convolved through the angular response of the 
instrument for each separate FOV.


The wide ($21^{\circ}$) FOVs [0,4,8] are large enough that this flux 
will be a combination of possibly several compact sources 
plus a strong contribution from the diffuse Galactic
continuum. The narrow ($3.7^{\circ} \times 21^{\circ}$)
FOVs [1-3, 5-7, 9-11] are
oriented relative to the Galactic Plane such that a compact 
source (or multiple sources) within one of these FOVs will 
dominate over the diffuse flux, and thus permit identification
of the compact sources and their intensities. The problem
is to determine the number of compact sources, their 
locations, and their fluxes --- as well as the diffuse
component distribution and flux.

The approach taken here is to assume that the possible
compact sources observed must be among the previously known 
hard X-ray sources in the Galactic Center region. Figure 1 
shows the 10 known sources used in this analysis. The hard 
X-ray sources, however, are highly variable --- any
combination of the sources could be contributing to the
observations; therefore, nothing is assumed about the source 
intensities except that they have to be positive.

These ten compact source locations were convolved 
through the angular response for each FOV to determine a 
response array for each source (Figure 3). The contribution 
of any source to the overall photon rates (Figure 2) is given 
by the response array for that source multiplied by the 
source flux. (The response arrays are actually energy 
dependent, but vary little below 200 keV where the main 
and narrow collimators are nearly opaque.) Each source has 
a unique response array, which allows deconvolution of the 
source contribution to the measured fluxes.


The assumed model for the Galactic diffuse continuum is 
that the flux comes entirely from the Galactic Plane, the 
scale height of this distribution is much less than
the $21^{\circ}$ 
FOV, and that the longitudinal distribution is flat for the 
range of observations here ($-35^{\circ} < l < 10^{\circ}$).
The resulting 
distribution is a line of constant flux along the Galactic 
Plane. The assumption of a constant longitudinal
distribution is based on previous observations of a Galactic ridge 
extending $\pm 60^{\circ}$ in longitude in the hard X-ray and soft 
gamma-ray bands
(e.g., \cite{whe76,wor82,koy89,geh93,str96,pur95,val98}),
and will be shown to be consistent with this observation as well.

The narrow Galactic Plane distribution was convolved 
through the angular response for each FOV to determine a 
response array for the diffuse component. This response 
array differs in units from the compact source array since it 
is expressed as the effective angular extent of the Galactic 
Plane through each FOV (Figure 3). The contribution of the 
diffuse flux to the overall photon rates (Figure 2) is given 
by this response array multiplied by the Galactic Plane flux, 
which is measured in [$ph/cm^{2}/s/rad$].

Given the set of 10 compact source and the Galactic diffuse 
response arrays, it remains only to find the combination of 
compact source (CS) and Galactic Plane diffuse (GP)
intensities which, when multiplied by the appropriate response 
arrays, best fit the observed rates in Figure 2. Statistically, 
with 12 data points and 11 intensities to fit, there is only one
remaining degree of freedom to judge the fit. Instead of 
trusting this fit, especially with the errors on these
observations, a more systematic and physical approach is applied.

The hard X-ray sources are highly variable, and often 
below the level of detection. Given this variability, the 
approach we have taken to determine the compact sources 
is to try an increasingly complex systematic combination of 
source models $GP+nCS$ ($n=0,1,...,10$), where we increase the 
number n of compact source combinations until a good fit is 
determined.

The first acceptable fits, which are also very good fits, 
occur for the combination $GP + 3CS$ model. There are actually
four combinations of three compact sources which give 
acceptable fits. All of these combinations have the sources 
GRO J1719-24 and 1E1740-2942 in common, but four 
choices of the third source give acceptable fits: GRO 
J1655-40 [$\chi^{2}_{\nu} = 5.40$, $\nu = 8$], OAO 1657-415 [5.55,8], GX 
340+0 [6.29,8], 4U1700-377 [9.95,8].

These four sources are located within $10^{\circ}$ of each other, and 
all lie along a line roughly parallel to the narrow collimator 
orientation (Figure 1), which explains why they could be 
confused in this analysis. While these measurements cannot 
absolutely determine which of the compact sources is the 
actual third source, there are two strong reasons to believe 
that this source is the marginally best-fit source, GRO 
J1655-40.

The first reason is that GRO J1655-40 was reported as 
undergoing a small flare during the period of this balloon 
flight, as determined by BATSE using occultation techniques
(\cite{har95}). The second reason is due to 
the spectral results below. This third compact source has a 
very hard spectrum, with a power-law ($\alpha = 2.3$) showing no 
sign of a spectral break below 330 keV. This hard spectrum 
is characteristic of black hole candidates. Of the four
candidate sources, OAO 1657-415, GX 340+0, and 4U1700-377 
are all neutron star binaries, while GRO J1655-40 is one of 
the strongest black hole candidates (\cite{bai95}). The 
spectrum supports the conclusion that GRO J1655-40 is the 
third compact source in these observations.


The best-fit combination of the response arrays for this 
model is shown in Figure 2. The overall fit is excellent 
[$\chi^{2}_{\nu} = 5.40$, $\nu = 8$], with the individual
flux rates given in 
Table 1. The uncertainties on these flux rates are determined
by varying the individual source intensities in the 
model fit independently until $\delta\chi^{2} = 1$.

\section{Spectral Decomposition}

To determine the individual spectra of these sources, one 
would like to perform the same analysis as above, only 
using smaller energy bands, extending the analysis from 30 
keV to 800 keV (the upper limit of good statistics), and 
assuming the four sources ($GP+3CS$) instead of deriving 
them. This analysis, however, tends to produce large
fluctuations, and hence spectral features, that are clearly
not real, an excess in one source spectrum having a corresponding 
drop in another. The measured spectra in the 12 FOVs are 
fairly smooth (Figure 4), exhibiting no obvious spectral
features. Therefore, it is natural to assume that the source
spectra themselves are smooth, that spectral features in
the sources are not combining in a way to produce overall 
smooth measured spectra.

In order to keep the source spectra smooth, we assume 
spectral models for the four sources, and then determine the 
best-fit parameters of these models. For the three compact 
sources we assume power-laws with exponential breaks:
\begin{equation}
f(E)=f(E_{o})(E/E_{o})^{-\alpha}e^{-(E-E_{o})/E_{break}},
(E_{low}<E<E_{high}),
\end{equation} 
where the spectral index $\alpha$,
exponential break $E_{break}$, and normalization $f(E_{o})$
are free parameters of the fit. This is a good general 
function because it can fit both a pure power-law 
($E_{break} > E_{high}$), or a thermal bremsstrahlung
($\alpha \simeq 1.3-1.5$).


For the diffuse continuum, we assume a single power-law 
plus positronium continuum, where the normalization of 
the two components is allowed to vary independently. 
There are two issues concerning this fit which should be 
addressed immediately. The first issue arises because of the 
assumed single power-law --- whereas several observations 
have shown a spectral break or steepening in the hard X-ray 
range (\cite{geh93,pur95}) which 
forcing a single power-law fit will not duplicate. The
ultimate test of this fit is the $\chi^{2}$. The fitting methods below 
turns out to be very robust for this data, in the sense that if a 
"wrong" fit is forced, the $\chi^{2}$ grows very quickly. We also 
confront this problem directly, however, by performing the 
spectral deconvolution twice, in the overall energy range 
(30-800 keV), and also in just the hard X-ray range (30-200 
keV), and confirming the consistency of the fit over the 
entire soft gamma-ray range.

The second issue concerns our use of a single incident 
positronium flux for all the FOVs of the instrument. While 
there has been evidence for a flat Galactic ridge for the 
power-law component, the same is not true for the positronium
continuum. Our deconvolution, therefore, fits the average
positronium continuum, not accounting for a gradient. 
This fit is justified, as will be shown, because our
measurements indicate that the gradient is less than our uncertainty 
for this component.

Using reasonable initial estimates of the spectral shapes and 
intensities, we apply a relaxation procedure to determine 
the best estimates of these spectra consistent with every 
FOV. An improved estimate of each source (compact or diffuse)
is obtained by subtracting all of the other estimated 
source fluxes convolved through the collimator response 
from each FOV, then finding the best-fit spectrum to the 
residuals consistent with the angular efficiency of the 
source in each FOV. This spectrum is used as the new best 
estimate for the source, and the procedure is then repeated 
on the next source. This procedure is iterated until the 
source fits have stabilized and the $\chi^{2}$'s are minimized. This 
technique is stable to initial estimates and order of
permutation. It is also sensitive to spectral models --- with 397 
degrees of freedom, spectral models that are forced (i.e. 
wrong shapes, parameters) quickly give unacceptable $\chi^{2}$'s.

The overall goodness of the fits are determined by convolving
the best-fit model spectra through the angular response 
of the instrument for each FOV, and finding the $\chi^{2}$
fit to the 
12 measured spectra simultaneously (Figure 4). Uncertainties
were determined by varying the parameters independently
until $\delta\chi^{2} = 1$. The spectral models fit the observations 
very well, with an overall
[$\chi^{2}_{\nu} = 392.1$, $\nu = 397$], which gives 
a 56\% probability that the four model spectra agree with the 
observations.


The best spectral fits to the compact sources are given in 
Table 2. 1E1740-2942 is the weakest of the sources, and 
shows a spectral break near or just below the lower energy 
range (30 keV) of our observations, which makes the spectral
index highly uncertain; therefore, it is assumed that the 
index fits a thermal bremsstrahlung ($\alpha = 1.4$). The spectra 
for these compact sources are all consistent with black hole 
candidates, and support the validity of the spectral
decomposition process.

The Galactic diffuse spectrum is well fit by a power-law 
with photon spectral index $\alpha = (1.80 \pm 0.25)$, $f(70 keV) = 
(1.95 \pm 0.28)\times10^{-4}$ $ph/cm^{2}/s/rad/keV$, and a positronium
continuum flux $(1.24 \pm 0.30)\times10^{-2}$ $ph/cm^{2}/s/rad$ (Figure 5).
The measured 511 keV flux from the Galactic Center is 
$(1.25 \pm 0.52)\times10^{-3}$ $ph/cm^{2}/s$, the analysis of which is
presented elsewhere (\cite{phd98}). This spectral index is
consistent with CGRO/COMPTEL measurements at higher 
energies (\cite{str96}), and RXTE measurements at 
lower energies (\cite{val98}).


\section{Self-Consistency Analysis}

In order to determine whether there is any evidence for a 
hard X-ray spectral break or steepening in the diffuse
continuum, the above analysis was repeated for just the hard
X-ray spectra (30-200 keV). If the Galactic diffuse does 
steepen in the hard X-ray range, then the power-law index 
for this fit should be significantly softer than the index for 
the entire energy range fit. The best-fit Galactic diffuse hard 
X-ray spectra (30-200 keV) has index $\alpha = (2.03 \pm 0.32)$ with 
$f(70 keV) = (1.89 \pm 0.28)\times10^{-4}$ $ph/cm^{2}/s/rad/keV$.
This fit is consistent with the
previous results over the entire range (30-800 keV), and
therefore these observations show no evidence for spectral 
steepening in the hard X-ray range.

Furthermore, this measurement strongly rules out the
best-fit spectrum determined by OSSE/CGRO observations in 
conjunction with SIGMA (\cite{pur95}). Figure 6 
shows the best-fit parameters and uncertainty curves for the 
(30-800 keV) fit. Shown for comparison are the parameters 
for the OSSE/CGRO hard X-ray results. As can be seen, this measurement
is inconsistent with the OSSE/CGRO results at 
$\delta\chi^{2} > 100$ (only taking into account the uncertainties
on this measurement), corresponding to a probability of consistency
of $<10^{-6}$.


As a check on the spectral deconvolution, the model fits are 
integrated over the range (30-200 keV) and the fluxes are compared 
to those determined from the source flux rates determined 
in Section 4. The results of these integrations are given in 
Table 1. The model spectra are consistent with the directly 
deconvolved hard X-ray fluxes for all four sources.

Given the best-fit spectral models, we can subtract the
compact source contribution from each FOV and determine 
whether the assumption of a flat Galactic ridge distribution, 
both for the power-law and the positronium continuum, is 
justified. The compact source spectra were convolved 
through the instrument response and subtracted from the 
flux in each FOV. Then, the remaining diffuse flux is
determined for each FOV separately. The flux rates in each
FOV can then be compared, and gradients determined.

Using this technique, the diffuse hard X-ray (30-200 keV) 
flux shows a gradient of $(1/f)(df/dl) = (-0.30 \pm 0.55)$ $rad^{-1}$ 
which is consistent with a flat distribution. For the
positronium flux, the gradient is
$(1/f)(df/dl) = (-0.52 \pm 1.19)$ $rad^{-1}$ 
which is also consistent with a flat distribution. Therefore, 
our assumption of a flat Galactic diffuse distribution is
self-consistent with the deconvolved source distribution and 
fluxes. These measurements place upper limits on the
gradient of the components of the diffuse continuum in the 
central radian of the Galaxy.

\section{Discussion}

From the measurements presented in this paper, we have 
concluded that the diffuse emission from the Galactic ridge 
can be well fit by a single power-law ($\alpha = 1.8$) plus the 
positronium continuum in the energy range 30-800 keV. 
The energy flux per logarithmic energy decade, $E^{2}f$, for our 
Galactic diffuse measurements is plotted in Figure 7 along 
with the COMPTEL/CGRO (\cite{str96}) and the 
RXTE (\cite{val98}) observations. From this 
plot we can see a smooth continuum (plus positronium) 
extending from 10 keV up to 30 MeV. However, our
normalization is slightly lower than the COMPTEL/CGRO and 
RXTE results, which is most likely due to our assumption 
that the latitude distribution of the diffuse emission is very 
narrow. For a broad latitude distribution, our overall
normalization could be low by factors as large as $\sim$2; however, 
the spectral shape is unaffected by this overall
normalization factor.

Also included is the bremsstrahlung and inverse Compton 
model of Strong et al. (1996), which we have smoothly
extrapolated from 10 keV down to 3 keV, and to which we
have added our measured positronium continuum, and the
thermal Raymond-Smith plasma component($kT = 2.9 keV$)
measured in the X-ray range by RXTE (\cite{val98}).
This model fits the
diffuse emission measurements well from $\sim$20 keV to 30 MeV. 
With the addition of the $\pi^{0}$ emission (\cite{ste88,ber94}),
this model fits the diffuse well up to $\sim$1 GeV 
(\cite{str96}). This model predicts that the HIREGS 
observations are dominated by the inverse Compton (plus 
positronium) emission, which is consistent with the large 
scale-height components measured by COMPTEL/CGRO 
and RXTE.


This model, however, underestimates the
diffuse continuum below 20 keV as measured by RXTE. 
While it is beyond the scope of this paper to determine the 
exact nature of this discrepancy, we present two possible 
explanations. One possibility is that thermal emission from 
a superhot ISM component could be contributing above 10 
keV. Such a component ($kT \simeq 7 keV$) was invoked to 
explain the hard component measured in ASCA observations
of the Galactic ridge in the energy range 0.7-10 keV 
(\cite{kan97}). Given the Galactic disk gravitational 
potential of $\sim 0.5 keV$ (\cite{tow89}), however, it is not 
clear how such a superhot component could be confined to 
the Galactic disk (e.g., \cite{kan97,val98}).

A more probable explanation is that the measured excess 
reflects the uncertainties in the cosmic-ray electron
spectrum below 100 MeV. The solar modulation of the
cosmic-ray electrons allows direct observation down to
only a few GeV (\cite{gol84,gol94,tai93}), and
synchrotron radio emission further constrains the electrons
down to 300 MeV (\cite{web83}). At lower energies, the
cosmic-ray electron spectrum is not constrained by observations. 
Theoretical calculations have been performed to extend the 
spectrum to lower energies (e.g., \cite{str95,ski93}),
which has in turn been used to estimate 
the diffuse gamma-ray spectrum. Inverse Compton scattering
emission below 20 keV would be predominantly produced
by electrons below 100 MeV, so we can turn the 
argument around to say that the measured hard X-ray spectrum
is constraining the cosmic-ray electron spectrum 
below 300 MeV. Therefore, we interpret the most likely 
cause of the discrepancy below 20 keV as due to theoretical 
underestimates of the true cosmic-ray electron spectrum at 
these previously unconstrained energies.

Our measurements are in strong disagreement with the 
coordinated OSSE/CGRO and SIGMA observations (\cite{pur95}), which 
show a steepening of the diffuse spectrum in the hard X-ray 
range ($\alpha = 2.7$). The OSSE/CGRO spectrum would require 
an anomalously large power ($\sim10^{43} erg/s$) to maintain the 
electron equilibrium against losses (\cite{ski96}).
One possibility is that the OSSE/SIGMA 
observations could still be contaminated by significant 
point source contributions.

Integrating the power-law component of the Galactic diffuse
and scaling for a rough model of the Galactic distribution
(\cite{ski93}) gives a total Galactic power 
output in the 30-800 keV band of $\sim6\times10^{37} erg/s$. Given an 
efficient emission process, i.e. inverse Compton scattering 
of interstellar radiation off of cosmic-ray electrons
accelerated in supernovae, this power is well within the total 
power injected to the Galaxy via supernovae, $\sim10^{42} erg/s$ 
(assuming $10^{51} erg$ every 30 yr).

\section{Conclusions}

HIREGS observations of the Galactic Center region, by 
combining wide and narrow FOVs simultaneously, allow 
separation of the Galactic diffuse continuum and the
compact source contributions, from ($-35^{\circ} < l < 10^{\circ}$).
The resulting diffuse soft gamma-ray spectrum is
found to be hard ($\alpha = 1.8$) 
down to 30 keV. The Galactic Plane distribution is found to 
be flat in this longitude range (Galactic ridge) with no sign 
of an edge, and upper limits are set on the gradients of the 
diffuse components in this region.

Comparison of the diffuse spectrum with the model of 
Strong et al. (1996) suggests that we are predominantly 
measuring the inverse Compton emission, in addition to the 
positronium continuum. The inverse Compton model is 
consistent with the large scale height components measured 
by COMPTEL/CGRO and RXTE in the neighboring energy 
bands, and with the smooth continuum measured from $\sim$20 
keV to 30 MeV. By minimizing the need to invoke
inefficient nonthermal bremsstrahlung emission in the hard
X-ray range, these measurements require only a small fraction
of the power provided by Galactic supernovae.
Discrepancies between the model and the 
measurements below 20 keV are most likely due to theoretical
underestimates of the cosmic-ray electron spectrum 
below 100 MeV; therefore, the diffuse hard X-ray/soft gamma-ray
spectrum is providing the first direct constraints on the cosmic-ray
electron spectrum below 300 MeV.

Given these conclusions, further measurements of the
diffuse hard X-ray/soft gamma-ray emission, both its spatial 
and spectral distributions, are crucial to separate the various 
components of the emission and better constrain the cosmic-ray
electron spectrum below 300 MeV. Furthermore, 
these conclusions suggest that nonthermal components 
extending from the hard X-ray into the soft X-ray bands 
should be analyzed in terms of constraining the cosmic-ray 
electron spectrum to even lower energies, dominated by 
inverse Compton scattering emission.

\acknowledgments

We are grateful to R. Ramaty for useful discussions, and
A. Valinia for providing the RXTE data.
This research was supported in part by NASA grant NAGW-3816.

\clearpage
 

\begin{table*}
\begin{flushleft}
\begin{tabular}{rcc}
{ } & Measured & Spectral Fit\\
\tableline
GP diffuse\tablenotemark{a} &$3.28 \pm 0.40$ &$2.63 \pm 0.36$ \\
GRO J1655-40\tablenotemark{b} &$3.19 \pm 0.31$ &$3.58 \pm 0.32$ \\
GRO J1719-224\tablenotemark{b} &$7.72 \pm 0.34$ &$8.23 \pm 0.43$ \\
1E1740-2942\tablenotemark{b} &$1.98 \pm 0.21$ &$1.72 \pm 0.21$ \\
\end{tabular}
\end{flushleft}
\tablenotetext{a}{[$10^{-2} ph/cm^{2}/s/rad$]}
\tablenotetext{b}{[$10^{-2} ph/cm^{2}/s$]}
\tablenum{1}
\caption{Source Flux [30-200 keV]. \label{tbl-1}}
\end{table*}

\clearpage


\begin{table*}
\begin{flushleft}
\begin{tabular}{rccc}
{ } & f(70 keV)\tablenotemark{a} & $\alpha$ & $E_{break}$ [keV]\\
\tableline
GRO J1655-40 &$2.72 \pm 0.22$ &$32.19 \pm 0.17$ &$>330$ \\
GRO J1719-224 &$5.86 \pm 0.22$ &$1.46 \pm 0.08$ &$44.3 \pm 2.9$ \\
1E1740-2942 &$0.99 \pm 0.10$ &$1.4$\tablenotemark{b} &$28.9 \pm 3.4$ \\
\end{tabular}
\end{flushleft}
\tablenotetext{a}{[$10^{-4} ph/cm^{2}/s/keV$]}
\tablenotetext{b}{defined (see text)}
\tablenum{2}
\caption{Compact Source Spectral Fits. \label{tbl-2}}
\end{table*}

\clearpage

\begin{figure}
\epsscale{0.8}
\plotone{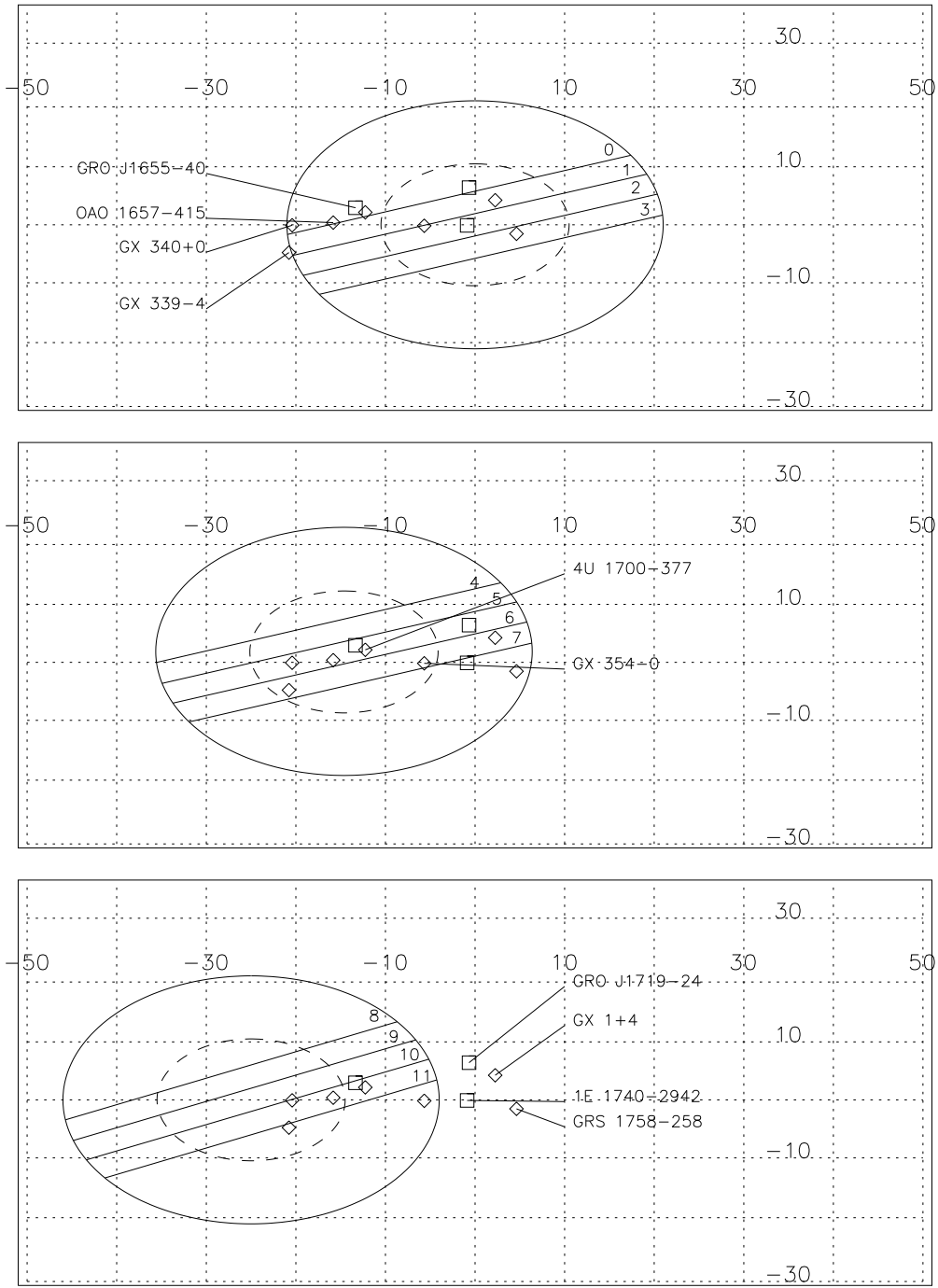}
\caption{The HIREGS observations in the Galactic 
Center and Plane ($-35^{\circ} < l < 10^{\circ}$). The 12 FOVs are 
numbered 0-11 for reference. Also shown are the 10 
trial compact sources used to deconvolve the data. 
The three compact sources (all black hole candidates) 
that this analysis shows were active are marked with 
squares.}
\end{figure}

\clearpage

\begin{figure}
\epsscale{1.0}
\plotone{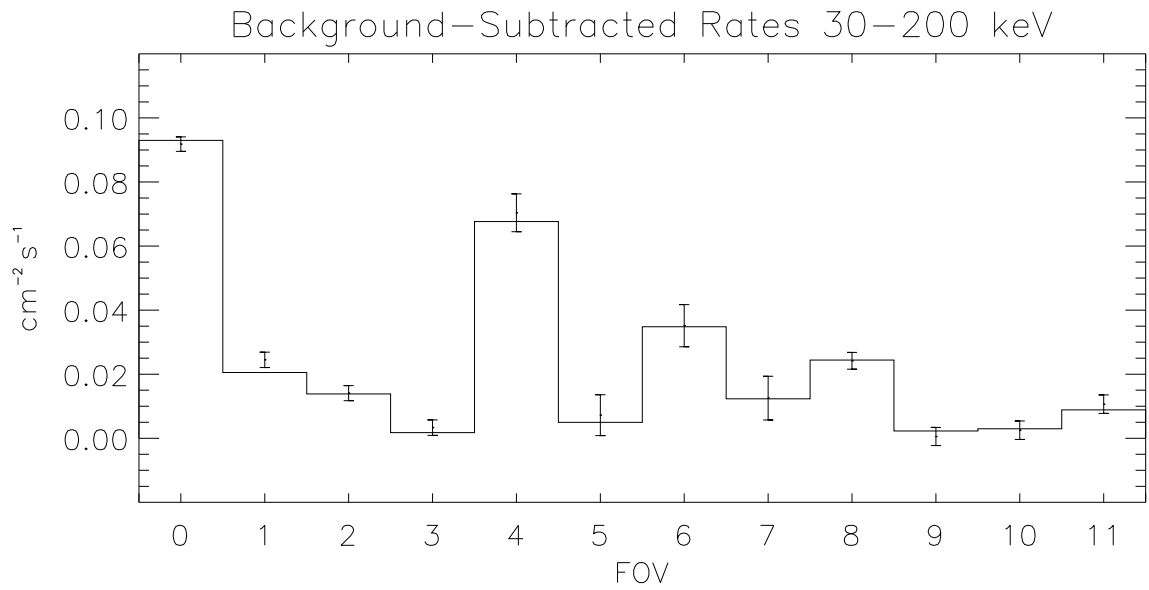}
\caption{The hard X-ray flux for the 12 FOVs shown in 
Figure 1. Also shown (solid line) is the best-fit fluxes 
for the Galactic diffuse, GRO J1719-24, 1E1740-2942, 
and GRO J1655-40 convolved through the angular 
response for each FOV [$\chi^{2}_{\nu} = 5.40$, $\nu = 8$].}
\end{figure}

\clearpage

\begin{figure}
\epsscale{0.8}
\plotone{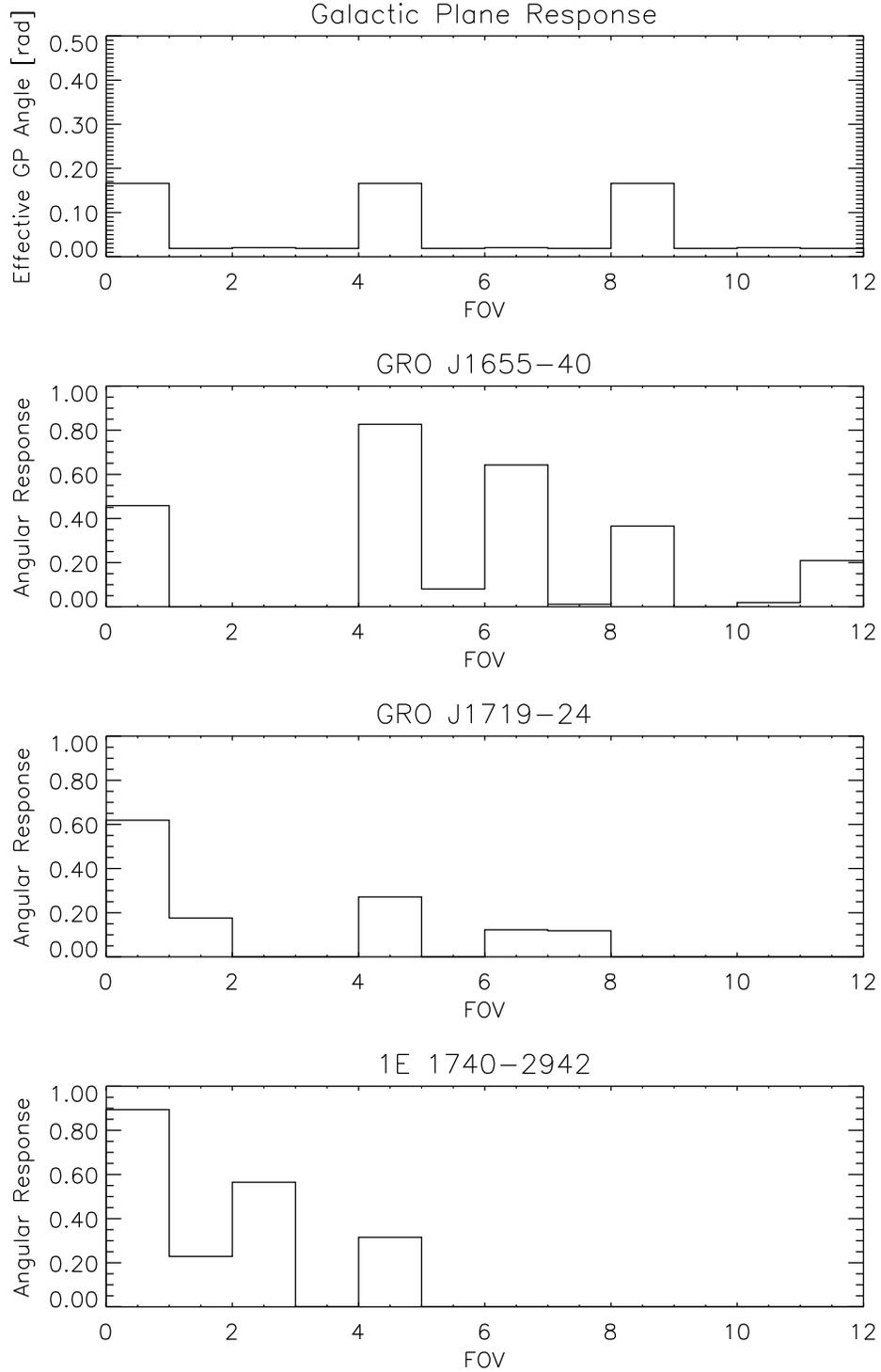}
\caption{The angular response for each FOV for the 3 
best-fit compact sources in Figure 1, and the Galactic 
Plane assuming a narrow ridge distribution. Each 
source has a unique response array, which allows 
deconvolution of the compact sources and the
Galactic diffuse flux.}
\end{figure}

\clearpage

\begin{figure}
\epsscale{0.8}
\plotone{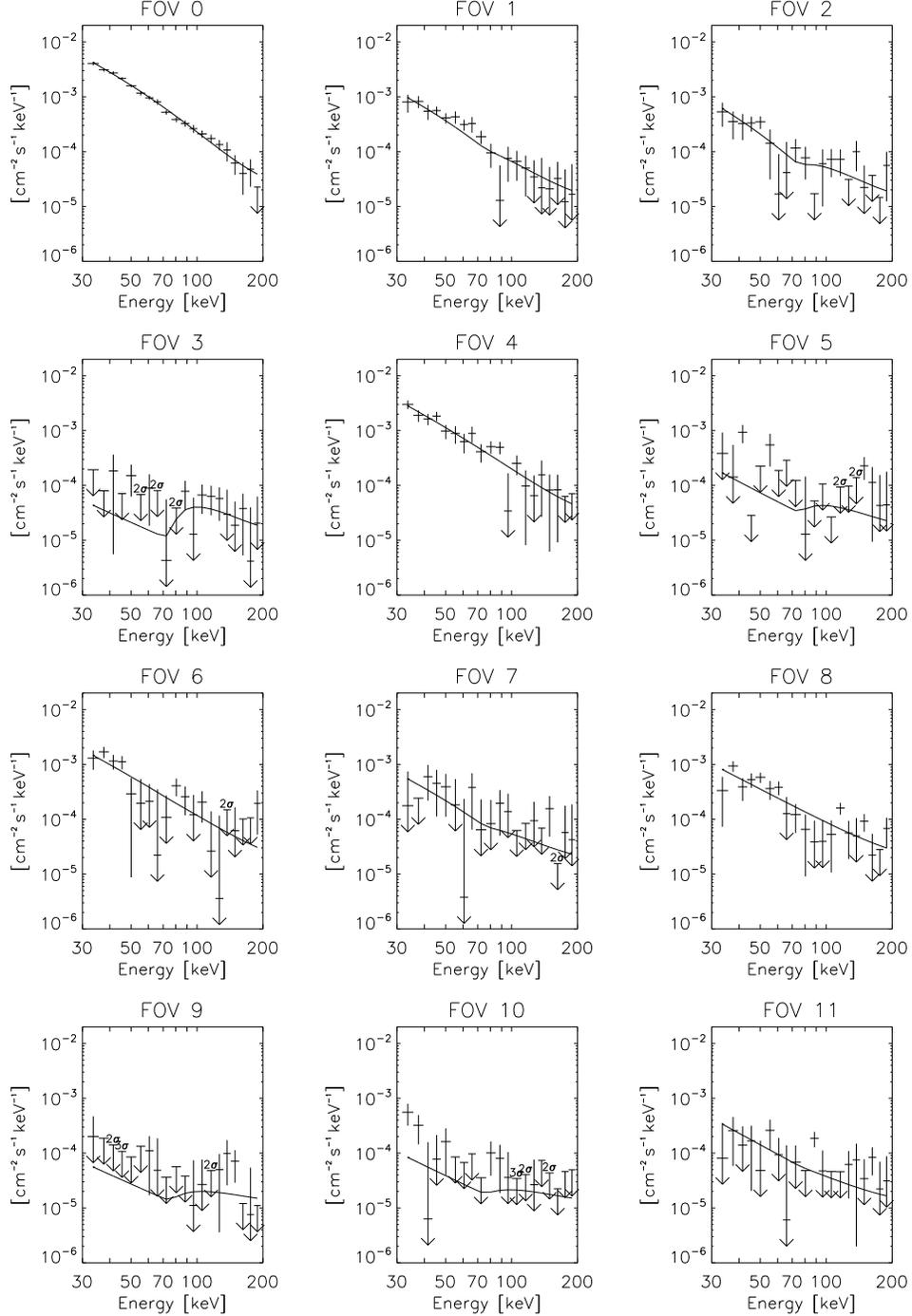}
\caption{The hard X-ray photon spectra for the 12
FOVs showns in Figure 1. Also shown (solid lines) are the
best-fit spectral models for the Galactic diffuse,
GRO J1719-24, 1E1740.7-2942, and GRO J1655-40
convolved through the angular response for each FOV
[$\chi^{2}_{\nu} = 392.1$, $\nu = 397$].
All limits are $1\sigma$ unless noted.}
\end{figure}

\clearpage

\begin{figure}
\epsscale{0.8}
\plotone{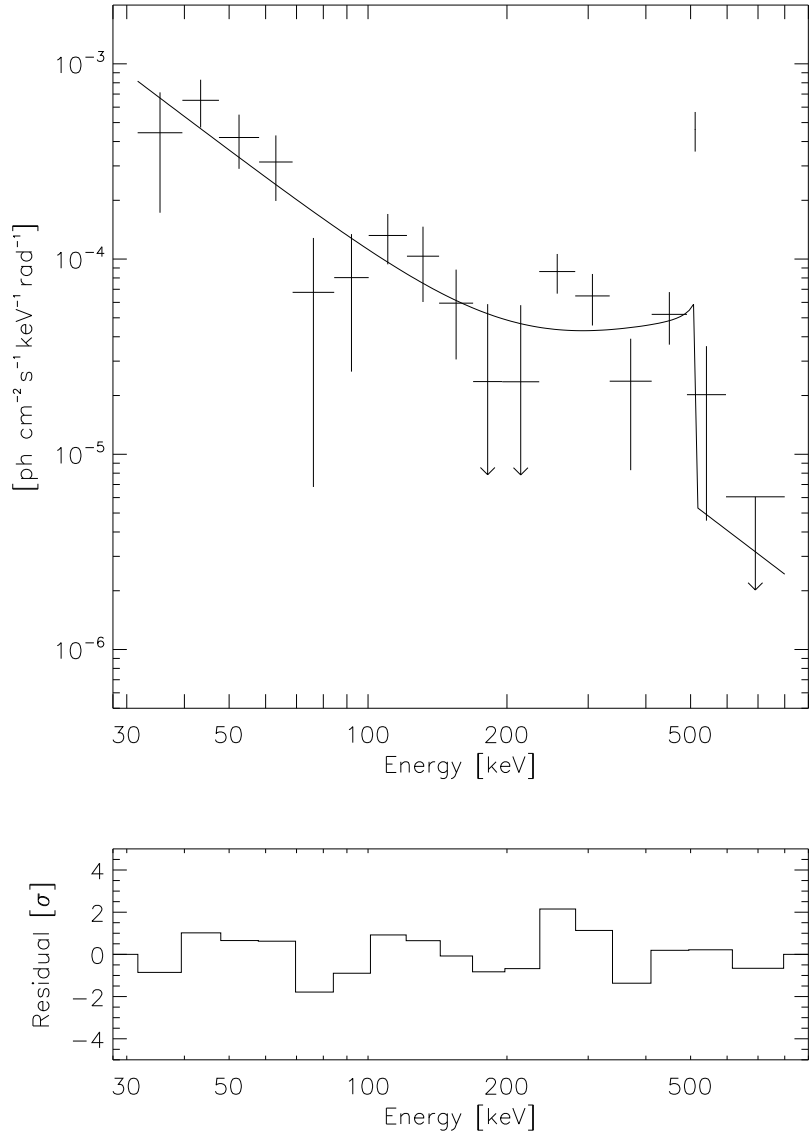}
\caption{The compact-source corrected Galactic
diffuse continuum. The best-fit spectrum (solid line) is 
shown. The 511 keV measurement is discussed in 
Boggs (1998).}
\end{figure}

\clearpage

\begin{figure}
\epsscale{1.0}
\plotone{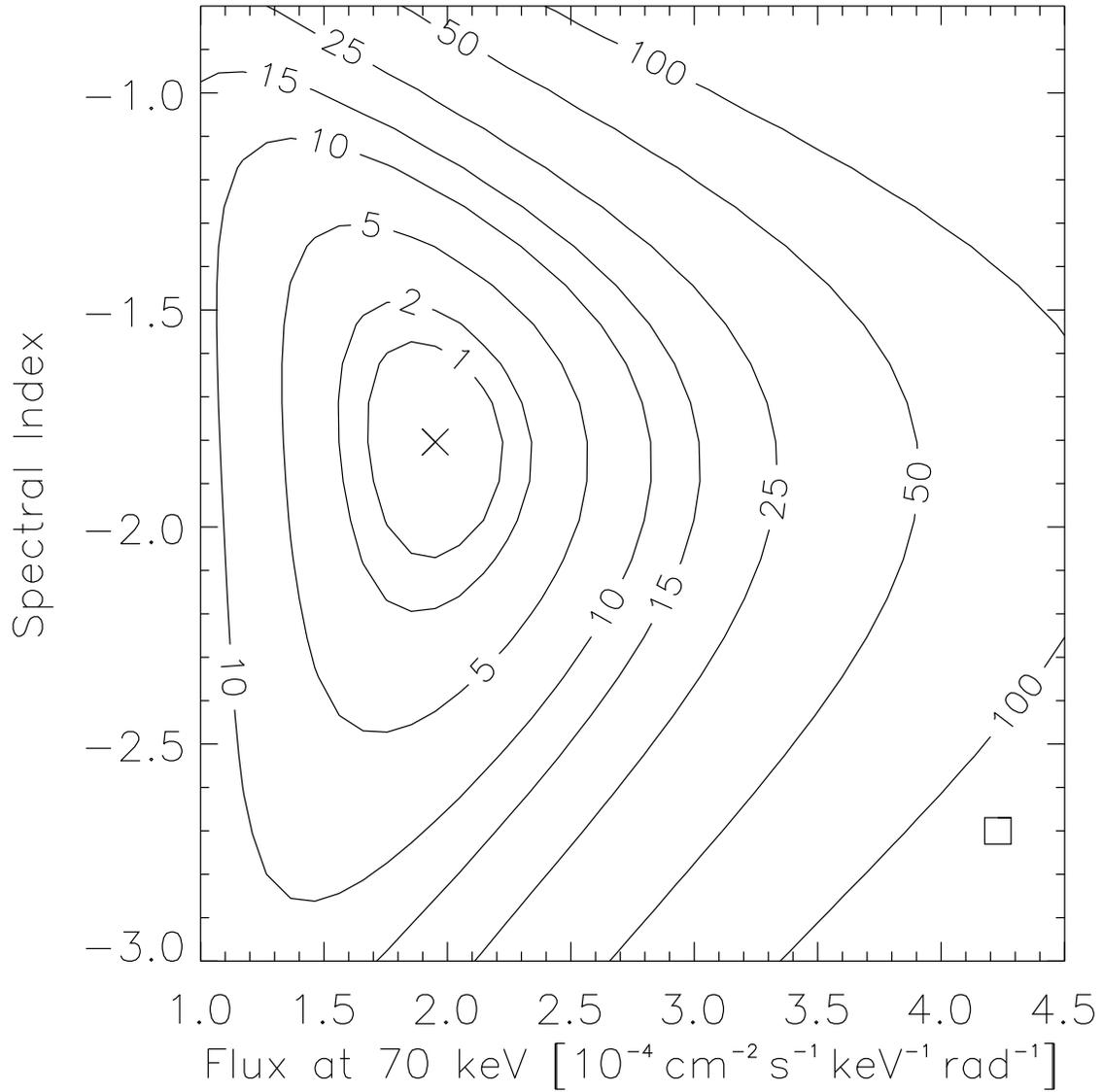}
\caption{The spectral parameter $\delta\chi^{2}$ curves for the 
(30-800 keV) diffuse continuum. Shown for comparison
(square) is the best-fit OSSE/CGRO hard X-ray spectral 
parameters (\cite{pur95}), which are inconsistent
with these measurements at the $\delta\chi^{2} > 100$ level.}
\end{figure}

\clearpage

\begin{figure}
\epsscale{1.0}
\plotone{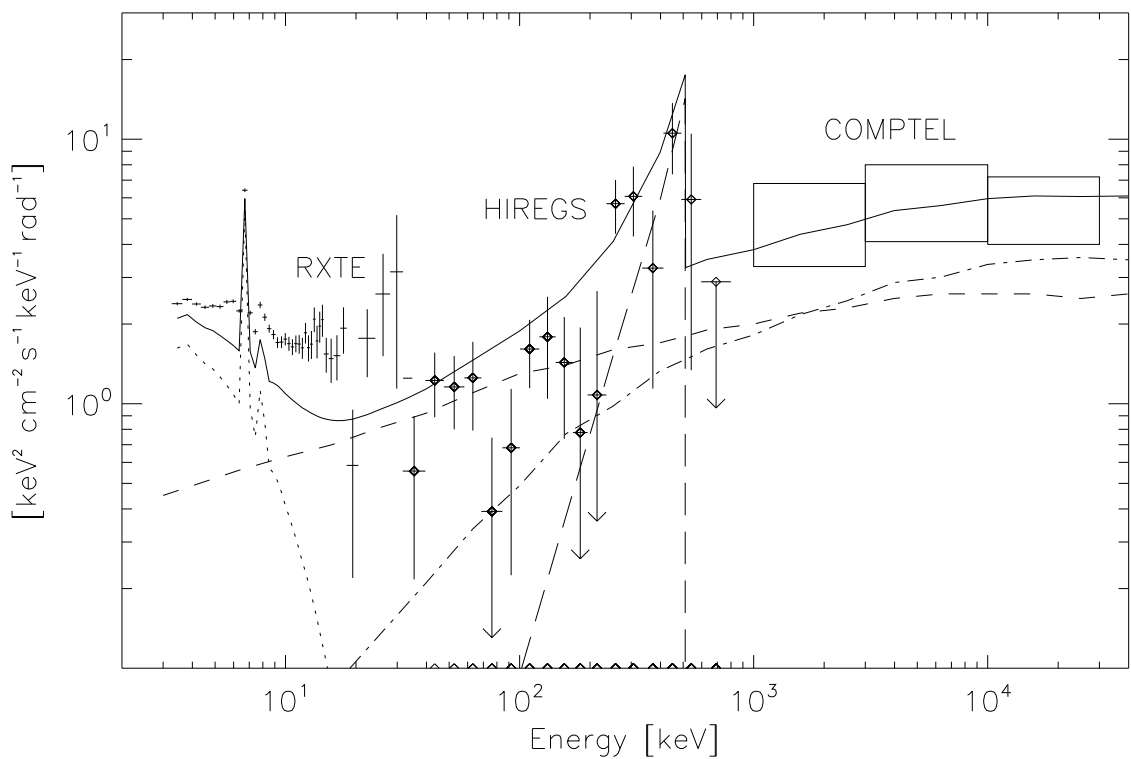}
\caption{The hard X-ray/soft gamma-ray
Galactic diffuse emission (RXTE - \cite{val98}, 
COMPTEL/CGRO - \cite{str96}). Also shown is 
the bremsstrahlung (dot-dashed line) and inverse Compton
scattering (short dashed line) 
model of Strong et al. (1996), with the HIREGS measured 
positronium continuum (long dashed line) and the
thermal Raymond-Smith plasma component(dotted line)
measured in the X-ray range by RXTE (\cite{val98})
added. The total of the model components is
also presented (solid line).}
\end{figure}


\clearpage

\begin{thebibliography}{}
\bibitem[Bailyn et al.\ 1995]{bai95} Bailyn, C. D. et al., 1995,
  Nature, 378, 157
\bibitem[Bertsch et al.\ 1994]{ber94} Bertsch, D., et al. 1994,
  \apj, 416, 587
\bibitem[Boggs 1998]{phd98} Boggs, S. E. 1998, Ph.D. Dissertation,
  University of California, Berkeley
\bibitem[Boggs et al.\ 1997]{bog97} Boggs, S. E., et al., 1997,
  The Transparent Universe, 2nd INTEGRAL Workshop, ESA SP-382, 153
\bibitem[Boggs et al.\ 1998]{bog98} Boggs, S. E., et al. 1998,
  Adv. Space Res., 21, 1015
\bibitem[Gehrels \& Tueller 1993]{geh93} Gehrels, N., \& Tueller, J. 1993,
  \apj, 407, 597
\bibitem[Golden et al.\ 1984]{gol84} Golden, R. L. et al., 1984,
  \apj, 287, 622
\bibitem[Golden et al.\ 1994]{gol94} Golden, R. L., et al. 1994,
  \apj, 436, 769
\bibitem[Harmon et al.\ 1995]{har95} Harmon, B. A., et al. 1995,
  IAU Circ., 6147
\bibitem[Harris et al.\ 1990]{har90} Harris, M. J., et al. 1990,
  \apj, 362, 135
\bibitem[Kaneda et al.\ 1998]{kan97} Kaneda, H., et al. 1997,
  \apj, 491, 638
\bibitem[Koyama et al.\ 1989]{koy89} Koyama, K., et al. 1989,
  Nature, 339, 603
\bibitem[Pelling et al.\ 1992]{pel92} Pelling, R. M., et al. 1992,
  SPIE, 1743, 408
\bibitem[Purcell et al.\ 1995]{pur95} Purcell, W. R., et al. 1995,
  Proc. 24th ICRC, 2, 211
\bibitem[Skibo 1997]{ski97} Skibo, J. G. 1997,
  The Transparent Universe, 2nd INTEGRAL Workshop, ESA SP-382, 555
\bibitem[Skibo \& Ramaty 1993]{ski93} Skibo, J. G., \& Ramaty, R. 1993,
  A\&AS, 97, 145
\bibitem[Skibo, Ramaty \& Purcell 1996]{ski96} Skibo, J. G., Ramaty, R.,
  \& Purcell, W. R. 1996, A\&AS, 120, 403
\bibitem[Stecker 1988]{ste88} Stecker, F. W. 1988, Cosmic Gamma Rays,
  Neutrinos and Related Astrophysics, eds. M. M. Shapiro \&
  J. P. Wefel (Dordecht: Reidel), 85
\bibitem[Strong et al.\ 1995]{str95} Strong, A. W., et al. 1995,
  Proc. 24th ICRC, 2, 234
\bibitem[Strong et al.\ 1996]{str96} Strong, A. W., et al. 1996,
  A\&AS, 120, 381
\bibitem[Taira et al.\ 1993]{tai93} Taira, T., et al. 1993,
  Proc. 23rd ICRC, 1, 333
\bibitem[Townes 1989]{tow89} Townes, C. H. 1989, IAU Symp., 136,
  The Center of the Galaxy, ed. M. Morris (Dordrecht: Kluwer), 1
\bibitem[Valinia \& Marshall 1998]{val98} Valinia, A., \&
  Marshall, F. E. 1998, \apj, 505, 134
\bibitem[Weber 1983]{web83} Webber, W. R. 1983,
  in 'Composition and Origin of Cosmic Rays,'
  Shapiro, M. M. (ed.), D. Reidel, 83
\bibitem[Wheaton 1976]{whe76} Wheaton, W. A. 1976,
  Ph.D. Dissertation, University of California, San Diego
\bibitem[Worrall et al.\ 1982]{wor82} Worrall, D. M.,
  et al. 1982, \apj, 255, 111
\bibitem[Yamasaki et al.\ 1996]{yam96} Yamasaki, N. Y.,
  et al. 1996, A\&AS, 120, 393
\end{thebibliography}
\end{document}